\documentclass{mn2e}
\usepackage{epsfig}
\usepackage{times}

\begin{document}

\title[KAT-7 observations of strong radio flaring by Circinus X-1] 
{A return to strong radio flaring by Circinus X-1 observed with the Karoo Array Telescope test array KAT-7}
 
\author[Armstrong et al.]{
{R.P. Armstrong$^{1,2}$, R.P. Fender$^{2,3}$, G.D. Nicolson$^{4}$, S. Ratcliffe$^{1}$, M. Linares$^{5}$, J.Horrell$^{1}$},\vspace{0.1in} \\
\emph{\LARGE{{L. Richter$^{1}$}, M. P. E. Schurch$^{2}$, M. Coriat$^{2,3}$, P. Woudt$^{2}$, J. Jonas$^{1,6}$, R. Booth$^{1,6}$, B. Fanaroff$^{1}$}}\\
       $^1$ SKA South Africa, 3rd Floor, The Park, Park Road, Pinelands, 7405, South Africa\\
       $^2$ Department of Astronomy, University of Cape Town, Private Bag X3, Rondebosch, 7701, South Africa\\
       $^3$ School of Physics and Astronomy, University of Southampton, High Field, SO17 1BJ, United Kingdom\\
       $^4$ Hartebeeshoek Radio Astronomy Observatory, PO Box 443, Krugersdorp 1740, South Africa\\
       $^5$ Instituto de Astrof\'isica de Canarias, c/ V\'ia L\'actea s/n, E-38205 La Laguna, Tenerife, Spain\\
       $^6$ Rhodes University, Lucas Avenue, Grahamstown, 6139, South Africa} 
\maketitle
\begin{abstract}

Circinus X-1 is a bright and highly variable X-ray binary which 
displays strong and rapid evolution in all wavebands. Radio flaring,
associated with the production of a relativistic jet, occurs 
periodically on a $\sim 17$-day timescale. A longer-term envelope
modulates the peak radio fluxes in flares, ranging from peaks in 
excess of a Jansky in the 1970s to an historic low of milliJanskys  during the years 1994 to 2007. Here we report first observations of this source 
with the MeerKAT test array, KAT-7, part of the
pathfinder development for the African dish component of the Square Kilometre
Array (SKA), demonstrating successful scientific operation for variable and transient sources with the test array. The KAT-7 observations at 1.9 GHz during the period 13 December 2011 to 16 January 2012 reveal in temporal detail the return to the Jansky-level events observed in the 1970s. We
compare these data to contemporaneous single-dish measurements at 4.8 and 8.5 GHz with the HartRAO 26-m telescope and X-ray monitoring from MAXI. We discuss whether the 
overall modulation and recent dramatic brightening is likely to be
due to an increase in the power of the jet due to changes in accretion rate or changing Doppler boosting
associated with a varying angle to the line of sight.
\end{abstract}
\begin{keywords} 
accretion: accretion discs -- ISM: jets and outflows --  instrumentation: interferometers -- radio continuum: stars -- stars: neutron -- X-rays: stars -- X-rays: binaries
\end{keywords}

\section{Introduction}

Circinus X-1 is a bright X-ray source with a 16.6 day periodicity
revealed in X-rays flares/dips (Kaluzenski et al. 1976) and radio
flaring (Haynes et al. 1978). This period probably corresponds to an
eccentric orbit in a binary system in which the accretion rate
increases dramatically around periastron passage.  In both radio and
X-ray bands Cir X-1 has shown strong evolution of its properties over
the half a century since its discovery. The 16.6-day modulation has
shown dramatic changes in overall intensity and shape in the X-ray
band (Saz-Parkinson et al. 2003; Linares et al. 2010a). Type I X-ray
bursts, identifying the accretor as a neutron star, were observed in
the 1980s (Tennant et al. 1986a,b) but then not again for a further 25
years (Linares et al. 2010b, 2010c, Papitto et al. 2010; Linares et al. 2010a). In a similar fashion, strong radio
flaring, reaching peak levels in excess of a Jansky at GHz frequencies, was
observed in the 1970s (Haynes et al. 1978, Nicolson, Feast and Glass 1980) but then declined
by a factor of $\geq 20$. Indeed semi-regular radio observations with
the Australia Telescope Compact Array (ATCA) for a 12-year period from
1994 onwards revealed no flares brighter than 50 mJy at in the 5-9 GHz
band (Tudose et al. 2008; Calvelo et al. 2012a). 

However, in the past few years Cir X-1 has begun to brighten in the
radio band. Nicolson (2007) reported peak radio flares reaching $\sim$
Jy levels at 8.5 GHz for the first time in 20 years, in observations
with the HartRAO 26-m telescope. Following the return to strong X-Ray flaring in June 2010 (Nakajima et al. 2010), Calvelo et al. (2010) reported a flare of $>500$ mJy at 8.5 GHz detected at HartRAO and subsequent flares of $170\pm20$ mJy at 5 GHz and $230\pm20$ mJy at 8 GHz observed with the
ATCA. Cir X-1 has returned to the strong radio flaring state last observed in the 1970s and early 1980s.

\begin{figure*}
\includegraphics[width=54mm]{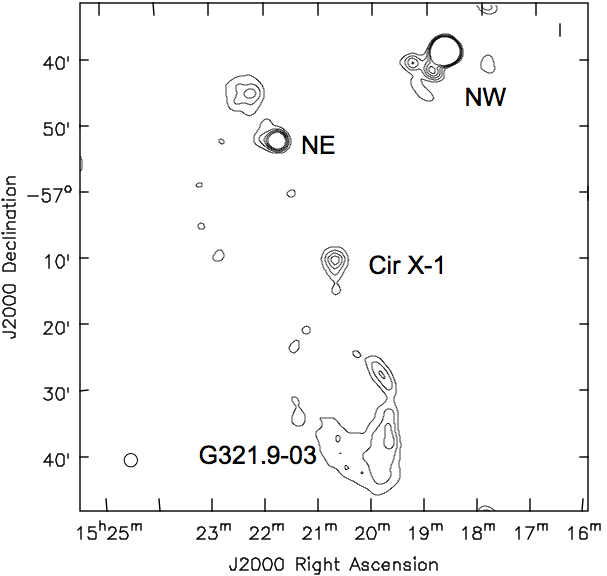}{Q}
\quad\includegraphics[width=54mm]{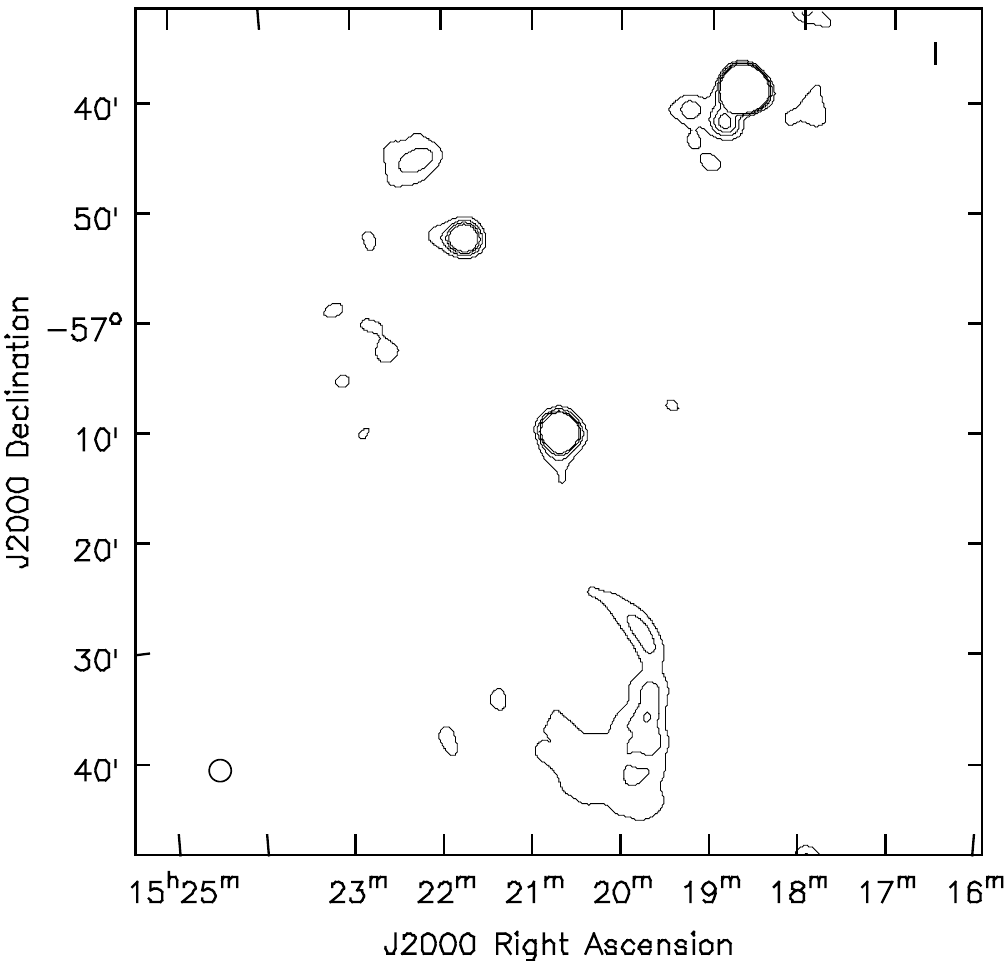}{F}
\quad\includegraphics[width=54mm]{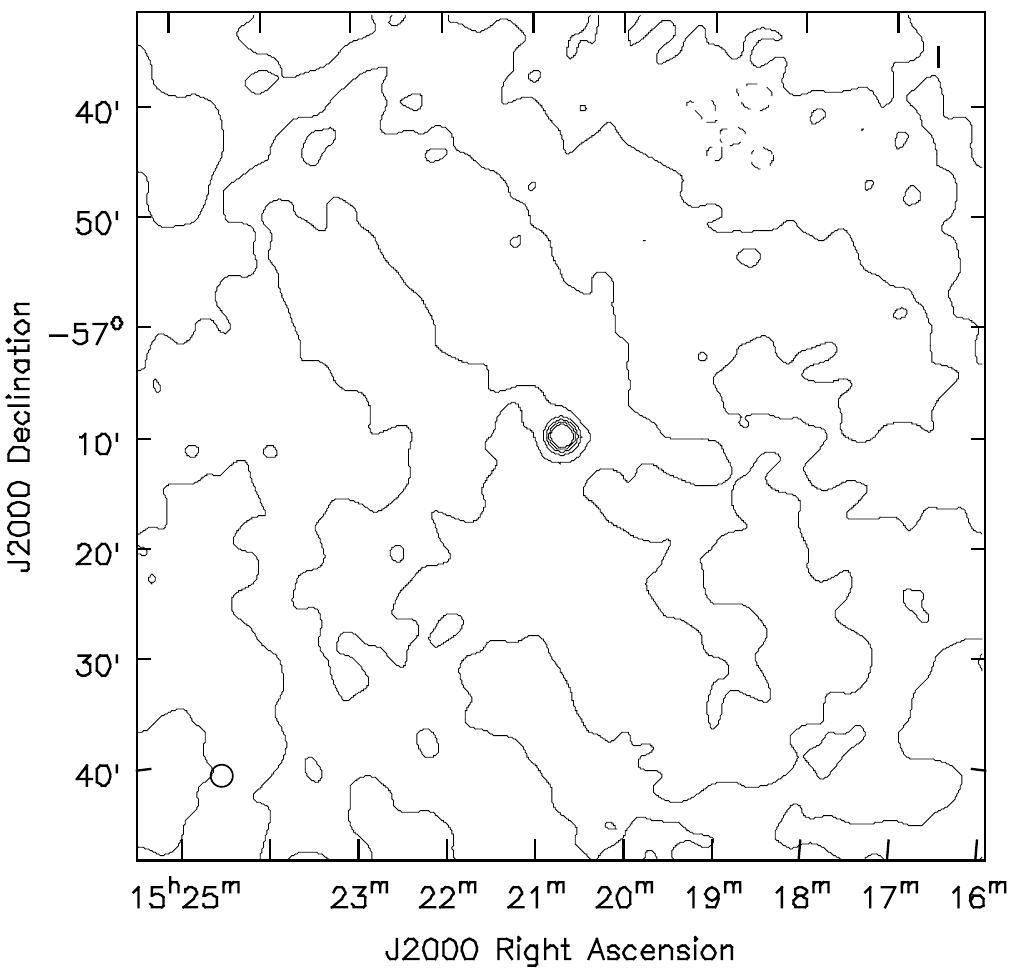}{F-Q}
\caption[] {
Variable radio emission from Cir X-1 in images from the KAT-7 interferometer. The leftmost panel shows a fiducial profile image obtained during the quiescent phase of the binary flare cycle during the period MJD 55922 - 55932. Cir X-1 is the bright (0.20 Jy) object in the centre of the field; at KAT-7 angular resolution the X-ray binary core is blended with the surrounding nebula. To the south the supernova remnant G321.9-0.3 can be clearly seen. NW is a strong source which lies beyond the half-power point of the primary beam. Source NE is used as a noise indicator in the estimation of measurement errors. The middle panel shows an image taken during the radio flaring phase (MJD 55919) for comparison; here Cir X-1 is flaring to 0.87 Jy. The rightmost panel shows the residual image produced after subtracting the fiducial profile from the flaring image. The only feature showing significant variability is Cir X-1; some apparent variability in extended structures away from the phase centre is apparent, such as G321.9-0.3, but this is beyond the half-power of the primary beam and of low signal-to-noise ratio. The contours in all the images are at (-0.2, 0, 0.2, 0.4, 0.6, 0.8) times the unit contour of 200 mJy, and North is up and East is to the left.
}
\label{im_quiescent}
\end{figure*}

What is the origin of this variable radio emission? In X-ray binaries
such flaring episodes are generally accepted to be synchrotron
emission associated with relativistic ejection events, in both black
hole (Fender, Belloni \& Gallo 2004; Fender, Homan \& Belloni 2009)
and neutron star (Migliari \& Fender 2006; Migliari, Miller-Jones \&
Russell 2012) systems. Fender et al. (1998) reported that on arcsec
scales the core radio source of Cir X-1 was resolved and oriented
towards larger-scale jet-like features seen in the surrounding radio
nebula, supporting the suggestion that the nebula was created by the
prolonged action of the jet (e.g. Stewart et al. 1991) similar to the
W50 nebula around SS 433 (Dubner et al. 1998; Tudose et
al. 2006). These radio structures have furthermore been found to have
X-ray counterparts on arcmin scales (Soleri et al. 2009; Sell et al. 2010)

More controversially, Fender et al. (2004) reported that sequential
brightenings in the arcsec-scale radio structure around Cir X-1
indicated an ultrarelativistic (bulk Lorentz factor $\Gamma \geq 10$)
flow from the core which was not directly observed. If correct, this
would indicate that a black hole is not required in order to
produce a highly relativistic jet. There is some evidence that similar
behaviour may have been observed in other neutron star XRBs (e.g. Fomalont,
Geldzahler \& Bradshaw 2001a,b). Recently, Cir
X-1 has been detected (Phillips et al. 2007, Moin et al. 2011) and resolved (Miller-Jones et al. 2012) at sub-arcsec scales using GHz VLBI and also via mm
observations with ATCA (Calvelo et al. 2012b). These
observations do not conclusively confirm or refute the
ultrarelativistic interpretation.

Radio astronomy is entering an age of rapid new development, driven in part by the push towards the Square Kilometre
Array (Carilli \& Rawlings 2004, Hall 2004). All of the new generation
of facilities (e.g. LOFAR, ASKAP, MeerKAT, MWA) have embraced the
study of transient and variable radio sources as key science. The
South African SKA precursor, MeerKAT, will be located in the Karoo
region of the Northern Cape. 
As part of the development of MeerKAT
(Booth et al. 2009), a scientific test array, KAT-7, has been
constructed and commissioned at the same site. Seven identical 12m-diameter radio dishes have a distribution optimized so as to have gaussian UV distribution, with highest weighting given to the optimization parameters of 4-hour tracks at 60 degrees declination (de Villiers 2007). Maximum baseline separation is 192m and minimum spacing is 24m. 
In this paper we report
observations of two flares of Cir X-1 with the KAT-7 array at 1.9 GHz, showing the potential of next-generation digital interferometers for transient science. 
In our analysis we combine these data with higher-frequency
measurements at 4.8 and 8.5 GHz made with the HartRAO 26-m telescope as
well as X-ray monitoring from MAXI.

\section{Observations and Analysis}
We report the first set of interferometric radio observations of Cir X-1 with
KAT-7, which occur during the period December 2011 to January 2012 
and span two flare cycles of the X-Ray binary. 

\begin{figure}

\includegraphics[width=85mm, ]{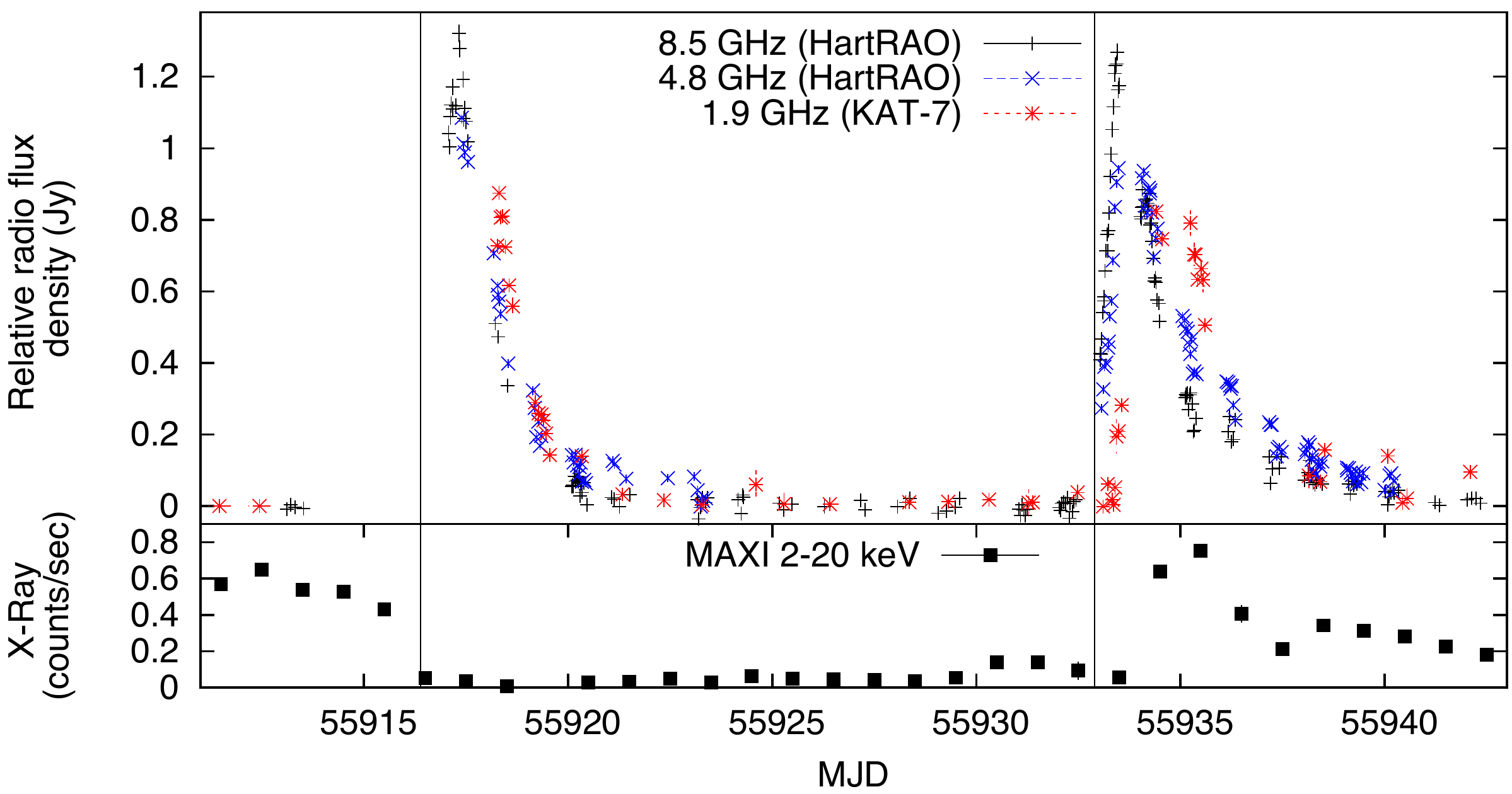}
\caption[] {Radio light curve of Cir X-1 obtained with KAT-7 (1.9 GHz,
  red crosses) and HartRAO (4.8 GHz, black circles and 8.5 GHz, blue stars). The
  system is clearly flaring to Jansky levels again. The vertical lines indicate the flare times predicted
  by the (updated) quadratic ephemeris of Nicolson (2007), which match well with
  the radio flaring. The X-ray behaviour is more irregular, with
  apparent orbit-long `on' and `off' phases. The vertical bars at each point represent the measurement noise (detailed in Section \ref{sec_kat} and \ref{sec_hart}), while the horizontal bars represent the observation length.}
\label{LC}
\end{figure}

\subsection{KAT-7 Observations at 1.882 GHz}
\label{sec_kat}
Cir X-1 observations were scheduled on KAT-7 as often as possible during a period spanning two flare cycles (MJD 55908-5545) and were planned to coincide with intensive observations with the 26-m single-dish telescope at HartRAO. Observation dates and lengths for KAT-7 are detailed in Table \ref{tab_log}. PKS 1934-638 was used as a flux density scale calibrator for the interferometric observations. A one minute observation of the phase
calibrator, PKS 1613-586, was interleaved every five minutes with
observations of the Cir X-1 field.
Data were recorded in two linear polarisations over a
256 MHz bandwidth centred at 1882 MHz, and correlated with an integration time of 1s. 

Longer observations ($>7$ hr.) were split into shorter segments of varying length (0.5 -- 2 hr.) after calibration depending on the data quality to obtain images with finer temporal
resolution. Less than 1\% of the channels in the central bandpass range needed to be discarded due to radio interference. 
These data were analysed in Stokes \emph{I} and polarisation properties ignored.

The data reduction for KAT-7 observations was performed in the {\sc CASA}\footnote{\emph{http://casa.nrao.edu/}} radio astronomy package as follows: a quiescent-state profile image of the field was obtained by combining the
longest observations during non-flaring parts of the flare
cycle (during MJD 55922-55932). This fiducial image is shown in Figure 1 (left panel, ``Q''). The profile contains imaging artefacts, distinct for each epoch due to varying numbers of operational antennas, as well as sidereal fluctuations primarily caused by a strong ($>3$ Jy) source to the north west of the field (``NW'' in Fig \ref{im_quiescent}), which moves through the steepest gradients of the primary beam as a function of frequency.

Observations from each epoch were then processed through an imaging `pipeline' (i.e. calibrated and imaged in an identical way). An example image produced in this way is shown in Fig 1, centre panel labelled ``F'' from MJD 55919.  The residual image of each observation was obtained by image-plane subtraction of the observations from the fiducial image (Figure 1, right panel, ``F$-$Q''). A conservative estimate of the measurement error was obtained by a combination in quadrature of the image noise and the apparent image fidelity, taken to be the variation in flux of the bright source to the north east (``NE'' in Fig \ref{im_quiescent}), which falls within the $\sim1^{o}$ primary beam of the KAT-7 dishes. Quadratic combination of these terms ensures that the contribution to the total measurement error by the fidelity term only becomes important when it is large enough to be of the order of the RMS image noise.

\subsection{HartRAO Observations at 4.8 and 8.5 GHz}
\label{sec_hart}

Contemporaneous observations were made at 4.8 and 8.5 GHz with the 26-m radio telescope at the Hartebeeshoek Radio Astronomy Observatory (HartRAO). The observations were part of a continuing monitoring program to study long term variability in Cir X-1 and to refine the ephemeris derived from onset times of the radio flares (Nicolson 2007). The current series of observations began in 2005, using Dicke-switched dual-beam radiometers at the above frequencies, with spanned bandwidths of 400 and 800 MHz respectively. The corresponding half-power beamwidths were 10 and 6 arc minutes and beam separations of 15 arc minutes were used at both frequencies.

The HartRAO data reduction strategy is a one-dimensional equivalent to that used for the KAT-7 data.  Drift scans were observed in both circular polarisations as part of the routine daily queue-scheduled observing on the 26-m telescope. When the source was predicted to be in a flaring state, based on the most recent ephemeris, twenty-four scans per day were scheduled at 8.5 GHz at half hourly intervals and 12 scans at 4.8 GHz at hourly intervals. Once the flare had decayed, or else 3-4 days after non-detection of a flare, 3-5 observations, uniformly spread out in hour angle, were scheduled each day at 8.5 GHz only. During non-flaring periods the scan responds to the Cir X-1 nebula and other nearby confusing sources, as well as strong sources in the sidelobes, resulting in a confusion-dominated scan which is repeatable from one cycle to another. 

Fiducial non-flaring scans were derived for both 4.8 and 8.5 GHz by averaging many scans taken during the non-flaring phases of the ~17 day cycle.  In the first stage of data analysis these fiducial scans were subtracted from the observed scans to remove the confused background. If the two polarised scans met specific data quality conditions they were added to reduce the noise. During times when there was no flaring activity the resultant scans were essentially flat with slow drift superimposed.  The drift was removed by fitting a second order polynomial to the scan but excluding the regions of the scan containing the telescope beams from a source centred on the scan. The final root-mean-square noise fluctuations were typically 40 mJy. Telescope beam patterns (one positive and one negative, spaced 15 arc minutes apart on the sky, are a consequence of the dual-beam Dicke switching) were then fitted at the expected position of the flaring source using a least squares algorithm to compute the flux density for each beam. The final flux density is the average of the two beams. When no flares were detected, the residual flux at 8.5 GHz was distributed about zero with an rms scatter of 15 mJy due to receiver noise. There is however a slow long term wandering of +/- 5 mJy in the mean residual flux, which is believed to originate in small uncorrected pointing errors.  When a flare is present the estimated flux error is increased by quadratically adding a flux dependant error proportional to 2.0\% of the measured flux density to allow for uncorrected pointing errors in declination. At 4.8 GHz the errors are smaller owing to the larger beam, the error due to noise being 12 mJy while the proportional error was 1.2\%.
\begin{figure}
\includegraphics[width=85mm, ]{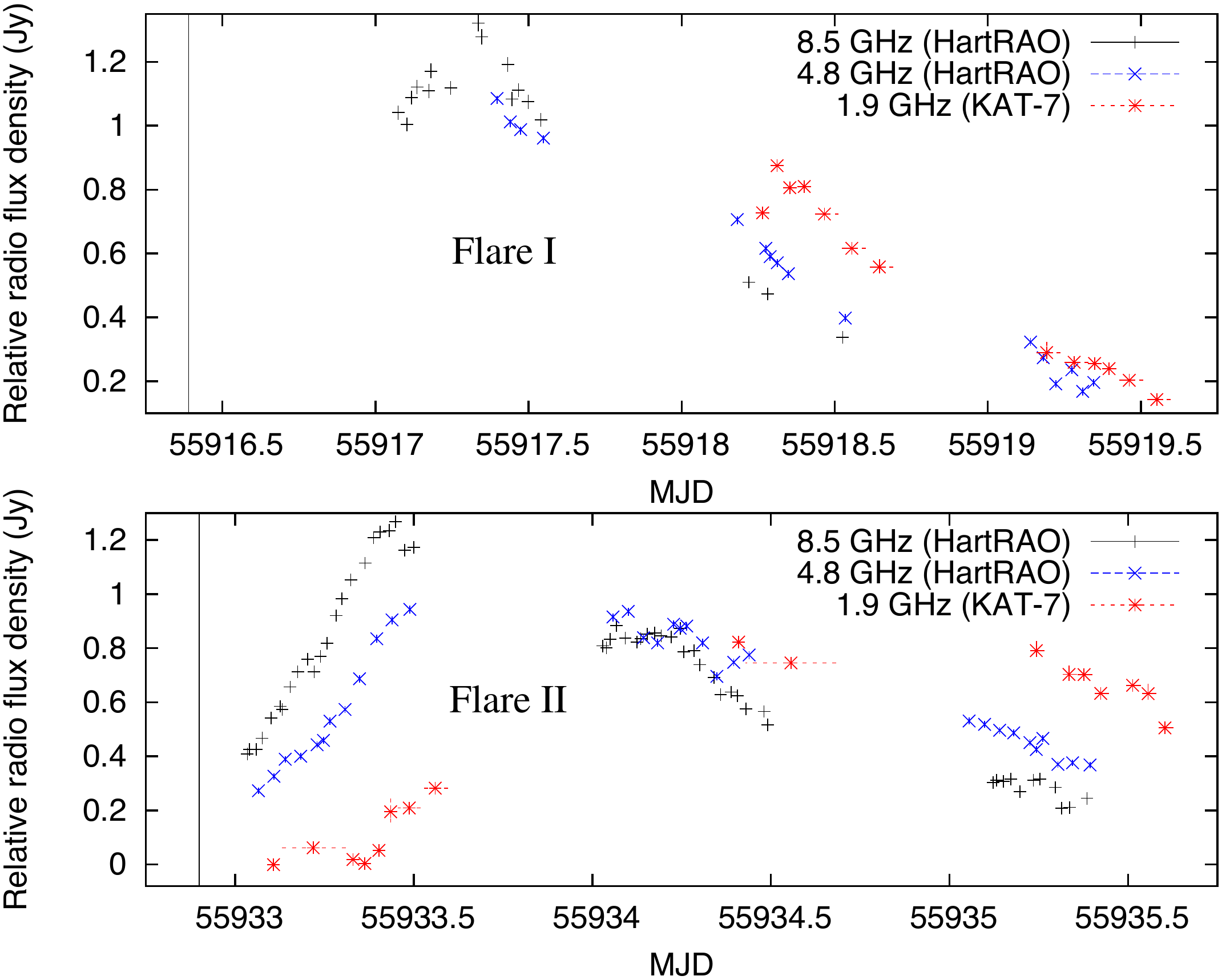}
\caption[]
{Analysis of the peaks of the observed radio flares. Measured data points from flare I (top) and flare II (bottom) are shown from the two radio telescopes in three frequency bands. The vertical lines demarcate the start of the flares according to Nicolson (2007). Once again, vertical lines on each point are error bars that represent the measurement noise, while horizontal lines represent the observation length. A linear interpolation to these data is used to determine the evolution of the spectral index between each of the frequency bands in Fig. \ref{fig_spix}.} 
\label{fig_lcz}
\end{figure}
\begin{table}
\scriptsize
\centering
\begin{tabular}{|c|c|c || c|c|c|}
Start Date & Duration& Antennas & Start Date & Duration & Antennas\\
(MJD) &(days)&used& (MJD)&(days)&used\\
\hline
55911.19&0.55&1234567 &55929.25&0.13&124567\\
55912.17&0.55&1234567 &55930.23&0.17&124567\\
55918.24&0.44&124567   &55931.30&0.17&124567\\
55919.15&0.52&124567   &55932.40&0.17&124567\\
55920.23&0.21&124567   &55933.09&0.58&124567\\
55921.28&0.10&124567   &55934.39&0.29&124567\\
55922.30&0.10&124567   &55935.22&0.42&124567\\
55923.23&0.10&124567   &55937.95&0.22&124567\\
55924.55&0.10&124567   &55938.38&0.22&24567\\
55925.24&0.11&124567   &55940.00&0.17&123567\\
55926.19&0.44&124567   &55940.40&0.20&124567\\
55928.28&0.17&124567   &55942.00&0.19&123567\\
\end{tabular}
\caption{Log of KAT-7 observations of Cir X-1. All dates are in MJD, and observation times in fraction of a day.}
\label{tab_log}
\end{table}

\subsection{Analysis}
Figure \ref{LC} shows the simultaneous radio light curve from KAT-7
(1.9 GHz) and HartRAO (4.8 GHz and 8.5 GHz). Plotted below in the sub-figure are the publicly-available soft X-ray fluxes (2-20 keV) from MAXI. 
The quadratic ephemeris of Nicolson (2007) predicts well the
times of radio flaring, although the X-ray behaviour is harder to
understand. The second radio flare (flare II) corresponds well to an X-ray flare,
but the previous event appears to correspond to a decline in the X-ray
flare and possibly alternating phases of `on' and `off' activity. 

\begin{figure}
\includegraphics[width=85mm, ]{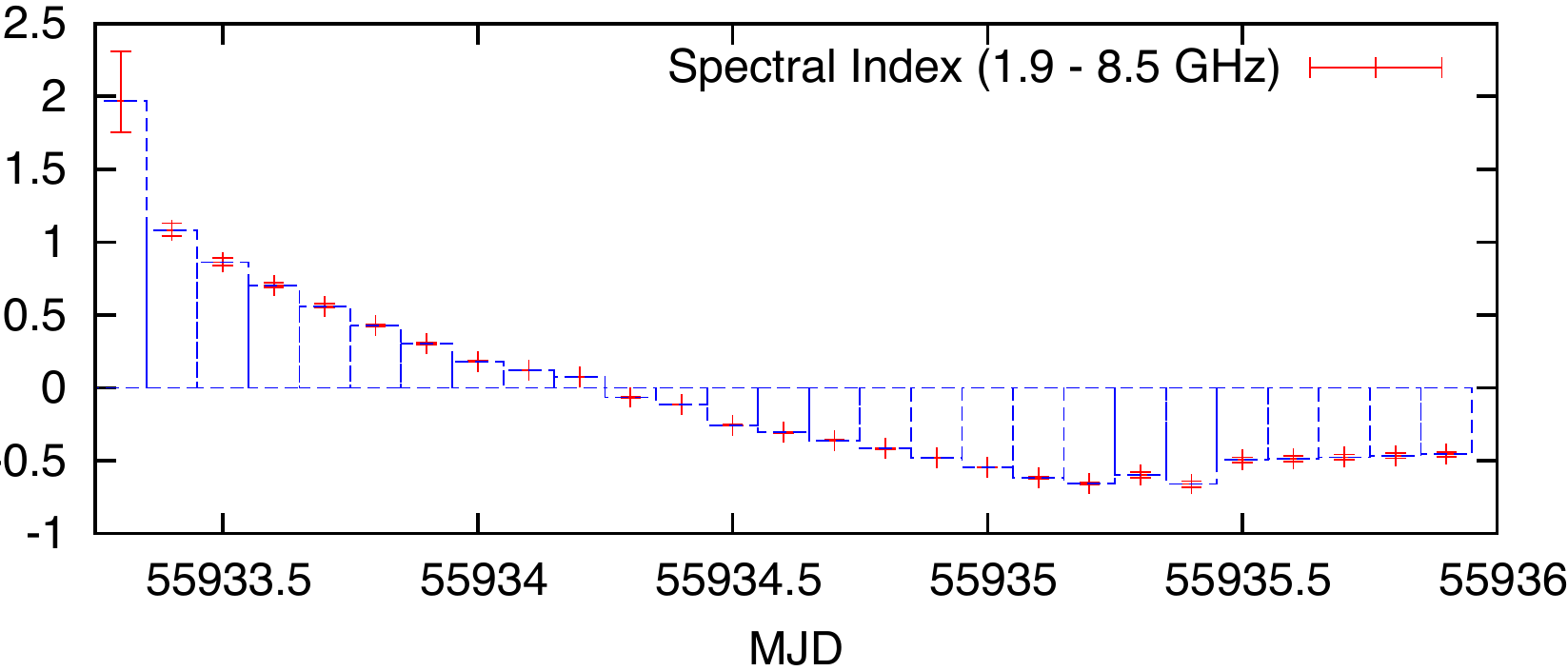}
\caption[]
{Evolution of the spectral index $\alpha$ (where S$_{\nu} \propto \nu^{\alpha}$) during flare II  between 1.9 and  8.5 GHz derived from a linear interpolation between measured points. The evolution from optically thick ({$S_{8.5} > S_{1.9}$)} to optically thin {($S_{1.9} >S_{8.5}$)} around the peak is clear, and is as expected since the earliest models of expanding radio source evolution (e.g. van der Laan 1966).}
\label{fig_spix}
\end{figure}

Figure \ref{fig_lcz} shows a close-up of the peak of the flares. During flare II, around MJD 55933, the 8.5 GHz radio flux data peak more energetically and earlier
than the 4.8 GHz data, which in turn lead the 1.9 GHz data. The 8.5 GHz data clearly go through a maximum at $\sim0.5$ days after the start of flare II, whereas both the 4.8 GHz and 1.9 GHz data still appear to be approaching their respective maxima at this time. This trend is also discernible in the first flare (flare I), but  here temporal coverage is sparser during the peak. The evolution of the spectral index $\alpha$ (where S$_{\nu} \propto \nu^{\alpha}$) during flare II is shown in Figure \ref{fig_spix}, clearly demonstrating the evolution from an optically thick to an optically thin phase.

This behaviour is commonly observed in
synchrotron-flaring sources (e.g. AGN, X-ray binaries) and explained by models
such as that of van der Laan (1966), which are based upon a simple,
expanding cloud of relativistic electrons and magnetic fields.  As the
cloud expands, the optical depth $\tau$ decreases at any given
frequency, and the flux rises until the source becomes optically thin
($\tau \leq 1$) at that frequency. Modulating this evolving optical
depth is an envelope of energy losses associated with adiabatic
expansion of the source into the surrounding medium, which results in
a dropping flux once the source is optically thin. Sources peak later
and less energetically at lower frequencies because the synchrotron optical depth
is higher at lower frequencies, and so by the time they become
optically thin the cloud has lost more energy to expansion. 

The 8.5 GHz and 1.9 GHz data also suggest multiple superposed outbursts during the larger flare events, particularly over the first three days after the start of flare II. 
At approximately day 2 after the start of flare II, there is what appears to be an optically thin brightening. This behaviour has similarities to the flare observed at the end of a sequence of flares from Cygnus X-3 by Fender et. al. 1997. It may be the result of a further jet-ISM shock, but one in which the local density is never large enough to make the source optically thick. The timescale of these events ($\sim$48 hours) is the same as previously reported for knot re-brightening (interpreted as episodes of turbulent interaction of the de-collimating jet with the local medium) of Cir X-1 (Fender 2004a). 

\begin{figure*}
\includegraphics[width=160mm]{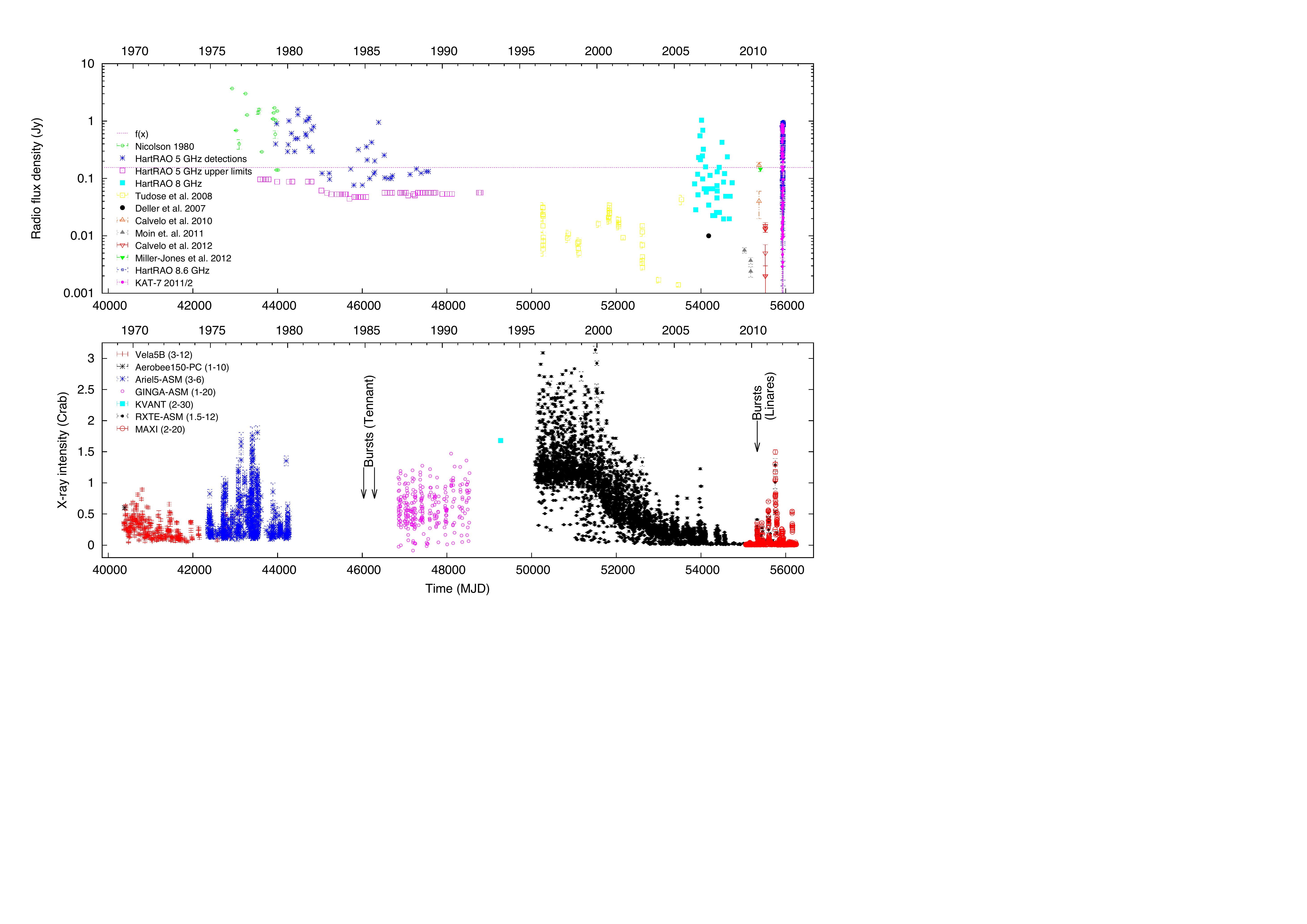}
\caption[]
{A long-term X-ray and radio light curve of Cir X-1 spanning 43 years of data taken by a number of X-ray missions and ground-based radio telescopes between 1969 and 2012. Also shown are the temporal locations of confirmed thermonuclear bursts (identifying Cir X-1 as a neutron star). The radio flares experience a $\sim20$-year lull from 1986-2006 during which no flares above 150 mJy (dotted line) are detected.} 
\label{fig_CirX1_longLC}
\end{figure*}

We have also compiled long-term X-ray and radio light curves of Cir X-1, in order to show the overall modulating envelope on strong radio flaring.  The X-ray light curve spans 43 years
of data taken by a number of X-ray missions, between 1969 and
2012. Apart from the data used by Parkinson et al. 2003 in a similar
compilation, we included X-ray intensity measurements from the
discovery observation (Margon et al. 1971), from the KVANT-TTM catalog
(Emelyanov et al. 2000) and from MAXI (Matsuoka et al. 2009). X-ray
intensities measured by Vela5B (Connor, Evans and Belian 1969), Ariel 5 (Smith and Courtier 1976),
RXTE-ASM (Levine et al. 1996) and MAXI (Matsuoka et al. 2009) were normalised
by the corresponding mission-averaged intensities from the Crab
nebula. GINGA-ASM data were taken from Parkinson et al. (2003). All
measurements use a soft X-ray band (somewhere between 1 and 30 keV),
yet each instrument operated in a slightly different energy band, as
indicated in Figure \ref{fig_CirX1_longLC}. The locations of confirmed type I (thermonuclear) bursts are also shown in the figure.

The long-term radio light-curve is compiled from an extensive set of previously-published radio measurements of Cir X-1, predominantly at $\sim$8.5GHz but also at $\sim$4.8GHz and $\sim$1.6GHz, between 1976 and 2012. Included are two previous compilations of shorter length. The first (Nicolson, Feast and Glass 1980) is of observations in the latter half of the 1970s, and shows Jansky-level flaring. The second compilation by Tudose et al. 2008 reports the extensive coverage by ATCA over the 12-year period from 1994-2006 showing no flares over 50 mJy during this period.

There is also new data from HartRAO from the period 1979-1992 ($\sim$MJD 44000 - 49000) showing the 5 GHz upper limits and 5 GHz radio detections , as well as 8.6 GHz radio detections around August 2006 ($\sim$MJD 54000). Also shown in the figure are observations by Deller et al. 2007 (see also Phillips et al. 2007), Calvelo et al. (2010), Moin et al. (2011), Calvelo et. al (2012) and Miller-Jones et al. (2012). Together, these data reveal a $\sim$20-year period beginning in 1986 during which there were detected no radio flares over 150 mJy, despite extensive campaigns with HartRAO and ATCA. This radio-quiet phase ceases in August 2006, and we show for the first time high-time-resolution coverage of strong radio flares during the 2006-onwards phase.

\section{Discussion}
These observations reveal two strong radio flares of Cir
X-1. Such events are considerably larger than those reported in the
1990s and 2000s by Fender (1996), Tudose et al. (2006), and Calvelo et
al. (2011), but are comparable to those observed in the 1970s (Haynes
et al. 1978, Nicolson, Feast and Glass 1980, Nicolson 2013a). They are the first high-time-resolution light curves of multiple flare events during the 2006-onwards strong-flaring epoch (this return to bright radio events previously noted by, e.g. Nicholson 2007, Deller 2007, etc.). Besides confirming that Cir X-1 exhibits expected behaviour for a synchrotron-flaring source, 
we also find a suggestion of superposed mini-flares, some of which occur when the source is optically thin. We suggest that they may be due to further jet-ISM shocks, and that they may be connected to `knot re-brightening', since they occur on the same timescale as the re-brightening of features\footnote{These `knot' features are not resolvable with KAT-7 or HartRAO.} in relativistic ejections during the low-radio-flaring epoch reported in Fender (2004a).

The long-term radio/X-ray light curve of Cir X-1 shows the return to a strong radio flaring phase from 2006 onwards, and a large variation in radio brightness with a $\sim$20-year low-flaring period. 
What mechanism is responsible for the large variation in radio luminosity over this long period? Two explanations are possible: that the large flares are either apparent (due, for example, to variable Doppler beaming) or intrinsic (possibly change in accretion rate, etc.). Both of these possibilities are discussed further below.

\subsection{Apparent Jet Variation}

It has been suggested several times (e.g. Fender et al. 2004a; Tudose et al. 2006; Calvelo et al. 2012; Miller-Jones et al. 2012) that the jets in Cir X-1 may have a small and varying angle to the line of sight. This has not been definitively established, nor is there any clear indication whether any angle changes might be random or more like the periodic precession of SS 433 (Margon 1984). Invoking regular precession to explain apparently bimodal behaviour of Cir X-1 seems improbable, since it would require the precession period to be extremely long
. It is still possible that the jet precesses aperiodically (which may explain what appear to be curved jets on all scales) but this would require it to have an aperiodic oscillation that occasionally brings us within the strongly-boosted line of sight.

It may be possible with more complete high-time-resolution coverage to test if the radio brightening is due to stronger relativistic boosting, since this would also produce more time contraction (blue shift) in the evolution of the approaching blobs. For identical physical conditions, the (relativistically moving) blob should appear both brighter and to evolve more quickly when viewed close to its axis of motion. What markers could we use to measure this? The time delay between the peaks at any two frequencies is entirely due to internal evolution in the model of van der Laan (1966) and might be a testable timescale with further observations of radio flares (i.e. the delay should be longer when the flares are fainter) if the jets are shown to remain broadly consistent between flares. 
If however, the jets are intrinsically different between flares, the finer angular resolution of VLBI observations would be required in testing if the jet variation is an apparent effect. Obtaining a VLBI image of the inner jet features during a strong flare episode would help verify this variable relativistic boosting scenario without relying on the assumption of constant jet properties of the unresolved high-time-resolution approach. 

\subsection{Intrinsic Variation}
Intrinsic changes in the Cir X-1 system could be responsible for the long-term jet variation. While the connection between radio flaring and X-ray emission is complex (see Fig 2, and also Soleri et al. 2009b), a relationship between, for example, the RXTE data and the onset of radio flaring in 2006 in Fig \ref{fig_CirX1_longLC} is evident.

Linares et al. (2010) favour changes in the accretion rate of Cir X-1 coupled with changes in the temperature of the neutron star as the most-likely explanation for the decrease in average X-ray intensity 
and the reappearance of the type I X-ray bursts respectively; such X-ray bursts seem to occur preferentially towards lower accretion rates --- high accretion rates suppress thermonuclear instabilities (e.g. Cornelisse et al. 2003, Linares et al. 2010) --- although this alone does not explain in this case why bursts were not observed between 2006 -- 2010, when the system had already faded in X-rays. Linares et al. (2010) suggest the cooling of the neutron star over this period as being the explanation for the observed behaviour (i.e. a thermal-hysteresis scenario; before Cir X-1 cooled, nuclear burning was stabilised and not bursty). 



In this scenario, the change in accretion rate responsible for the X-ray behaviour (Linares et al. 2010) may also be related to the generation of more-powerful, bursty flows from the neutron star, and hence to the observed stronger synchrotron radio emission during the 2006-onwards era. From the long-term data, this seems to be a possible and likely scenario; as the average and peak X-ray intensity decreased sufficiently from 1996 -- 2006, strong radio flaring switched on, and Jy-level radio flares are again observed. Some time later, type I X-ray bursts are again observed, seemingly unconnected to the radio behaviour;  while type I X-ray bursts may have a thermal hysteresis dependence, strong radio flaring is more likely only dependent on accretion rate.

\section{Conclusions}
Circinus X-1 continues to reveal new aspects of its behaviour, and is arguably the best laboratory for relativistic jet astrophysics in the southern hemisphere, providing as it does regularly spaced and predictable events during its periastron passage cycle. It is furthermore an excellent control to the large population of jets associated with accreting black holes. 

The observations reported here with KAT-7 are the first to be published with time resolution fine enough to resolve the individual flare features since its return to the radio-bright state first observed in the 1970s that disappeared for two decades. That these scientific observations were made with the seven-element test interferometer for MeerKAT, itself a precursor to the Square Kilometre Array, illustrates the wide range of science that can be done on variable and transient radio sources on the path to the SKA.

Cir X-1 continues to reveal further clues on the nature of accretion and relativistic jet formation around neutron stars. The long-term radio/X-ray light curve provides evidence that the reappearance of thermonuclear bursts from Cir X-1 are most likely not directly connected the return to strong radio flaring. Of the two scenarios that may be responsible for long-term radio brightness variation and recent dramatic brightening -- either an increase in the power of the jet due to a decrease  in accretion rate or changing Doppler boosting associated with a varying angle to the line of sight -- we suggest that the former is more likely.


\section*{Acknowlegements}
This work was carried out as an early science project during the
commissioning of the KAT-7 telescope. We thank the SKA South Africa project for the
time allocated to Cir X-1 observations and the engineering teams for
the smooth operation of the telescope.

This work made use of data supplied by the ASM/RXTE team, by RIKEN,
JAXA and the MAXI team, by the High Energy Astrophysics Science
Archive Research Center (HEASARC) at NASA's Goddard Space Flight
Center, and by the Hartebeeshoek Radio Astronomy Observatory.

RPA, MPES and MC acknowledge the financial assistance of the National Research Foundation (NRF) towards this research.  RPF is supported in part by European Research Council Advanced Grant 267697 ``4 pi sky: Extreme Astrophysics with Revolutionary
Radio Telescopes''. ML thanks M. Kachelrie{\ss} and the Physics department at the Norwegian University of Science and Technology for their hospitality. MPES is funded through the Claude Leon Foundation Postdoctoral Fellowship program.


%

\end{document}